\documentclass[trackchanges,twocolumn]{aastex701}

\usepackage{amsmath}	
\usepackage{enumitem}
\usepackage{verbatim} 
\usepackage[caption=false]{subfig}
\usepackage{array}

\usepackage{listings}
\usepackage{hyperref}
\renewcommand{\added}[1]{#1}
\begin{document}

\title[]{Dynamical stability of double planets in compact multiplanetary systems}

\makeatletter\let\frontmatter@title@above=\relax

\newcommand{\todo}[1]{\textcolor{red}{ToDo: #1}}

\author[0000-0002-8658-3811]{Wolf Cukier}
\email{wcukier@uchicago.edu}
\affiliation{Department of Astronomy \& Astrophysics, University of Chicago, 5640 South Ellis Avenue, Chicago, IL 60637, USA}

\author[0000-0002-6939-9211]{Tansu Daylan}
\email{tansu@wustl.edu}
\affiliation{Department of Physics and McDonnell Center for the Space Sciences, Washington University, St. Louis, MO 63130, USA}



\begin{abstract}

The ``peas-in-a-pod'' architecture of planetary systems, where planets are roughly evenly sized and log-spaced, is pervasive throughout known exoplanetary systems.  However, many systems deviate slightly from this trend in that they have a ``gap'' where no planet has been detected, where one might have otherwise expected one to exist.  There are many common explanations for the existence of these gaps, including the existence of as-yet-undiscovered planets lurking in these gaps; however, we propose that one potential explanation for some of these gaps is that they are filled by binary planets, a pair of gravitationally bound planets orbiting a common center of mass outside of both bodies.  Specifically, in this work, we investigate the dynamical stability of these binary planets as compared to the more conventionally assumed single planets using Monte Carlo N-body simulations.  We find that binary planets are slightly less stable than single planets.  Overall, we find that binary planets and their cousins in exomoons provide a plausible explanation, at least from a stability perspective, for the existence of some dynamical gaps.

\end{abstract}


\section{Introduction}
\label{sect:Introduction}

One of the outstanding results from the Kepler and TESS missions has been the observational evidence for dynamical gaps in multiplanetary systems \citep{Lissauer2011, Fabrycky2014, Jontof-Hutter2014Kepler79sLowDensity, Rice2024OrbitalGeometriesStellar}. At the same time, a large number of exoplanetary systems exhibit a ``peas-in-a-pod'' architecture \citep{Weiss2018}, with planets of roughly equal radii roughly evenly log-spaced in semi-major axis \citep{Millholland2017, Weiss2018}. There are also exoplanetary systems that violate this regular spacing and instead have gaps where one might have otherwise expected a planet to be \citep{Daylan2021a}. There have been several proposed explanations for these gaps, either through explaining the simple absence of a planet, potentially due to outer giant planets \citep{HeWeiss2023} or through explaining the existence of a non-transiting planet, for example due to it having a high inclination \citep{Becker2016} or by it being small enough that it evades detection by our telescopes \citep{Lammers2025GapGiantAssociationAre}.

In this paper, we propose binary planets---system sof two mutually bound planets with a barycenter outside their physical radii \citep{Stern2002RegardingCriteriaPlanethood}---as a potential explanation for a fraction of observed gaps and investigate the stability of binary planets within compact multiplanetary systems with dynamical gaps. The transits of a binary planet pair would exhibit significant variations in transit duration, transit timing, and transit shapes, which could lead to such objects being flagged as false positives \citep{Chakraborty2022TransitDurationTiming, Lewis2015EXTRASOLARBINARYPLANETS} or not being flagged at all. Although there are no known exo-binary planets \added{orbiting a star} to date, the Pluto-Charon system in our Solar System satisfies this definition \added{(or the analogous definition for binary dwarf-planets)}, the Earth-Moon system comes close, \added{and it has been hypothesized that Neptune's moon Triton was once a member of a binary \citep{Agnor2006NeptunesCaptureIts}}. \added{Free-floating planetary-mass binaries such as 2MASS J11193254-1137466 \citep{Best2017YoungDwarf2MASS} also satisfy our definition, although they likely form in a distinct manner from binary planets orbiting stars \citep{Huang2024DynamicsBinaryPlanets, Perets2012OriginPlanetsVery}.} Additionally exomoons, such as the proposed candidates Kepler 1625 b-i \citep{Teachey2018}, Kepler 1708 b-i \citep{Kipping2022ExomoonSurvey70}, and DH Tau Bb \citep{Lazzoni2020SearchDisksPlanetary}, inhabit an incredibly similar dynamical regime, differing only in mass ratio and separation.   Investigations of binary planets and exomoons are thus highly complementary. Irregular satellites of giant planets in the Solar System serve as analogs for the capture mechanisms that explain binary planets, offering valuable dynamical laboratories \citep{Gaspar2011}.


Interpreting the dynamical gap between Mars and Jupiter, Kepler had stated: ``Between Mars and Jupiter, I place a planet'' in his \textit{Mysterium Cosmographicum} (Cosmic Mystery; 1596). While we recognize today the invalidity of what would eventually become the Titius-Bode law, it is reasonable to expect that some apparent dynamical gaps may be due to massive objects occupying them but otherwise not observationally accessible.  Given that several studies hypothesize the existence of ordinary single planets in these dynamical gaps \citep{Lammers2025GapGiantAssociationAre}, we believe it is worthwhile to consider the existence of other potential bodies which, while less likely to form, are more likely to exhibit orbital parameters that make them observationally inaccessible, at least to conventional exoplanet detection implementations. In general, directly forming a binary in a compact multiplanetary system without disrupting the system appears to be challenging. However, several proposed formation mechanisms exist for binary planets, or their younger cousins, exomoons, such as formation through giant impacts \citep{Gayon2008AreRetrogradeResonances}, similar to proposed formation mechanisms of our Moon \citep{Hartmann1975SatellitesizedPlanetesimalsLunar, Cameron1976OriginMoon} and Charon \citep{Woolfson1999NeptuneTritonPlutoSystem, Canup2005GiantImpactOrigin, Canup2010GIANTIMPACTORIGIN}, pebble accretion of the two planets simultaneously \citep{Konijn2023FormingEqualMass}, and capture through tidal dissipation during planet-planet scattering \citep{Ochiai2014EXTRASOLARBINARYPLANETS, Lazzoni2024BinaryPlanetFormation}.  

To test this hypothesis, we investigate the dynamical stability of this proposed solution of binary planets filling dynamical gaps.  While work has been done that has studied the stability of binary planets in 3-body systems, binary planets or exomoons orbiting a host star \citep{Vilke2001EvolutionMotionBinary} and even binary planets or exomoons orbiting a star with a 4th body as a large perturber \citep{Quarles2021ExomoonsSystemsStrong}, the stability of binary planets and exomoons in compact multi-planetary systems has not yet been investigated.


The rest of the paper is as follows. In Section \ref{section: stability}, we describe our N-body simulations of compact multi-planetary systems; in Section \ref{sec: results}, we present the stability of binary planets in our multi-planetary systems and the regions of parameter space that allow such stability.  Finally, the implications of our results, along with caveats from modeling assumptions, are presented in Section \ref{sec: conclude}.

\section{Stability in a multiplanetary system}
\label{section: stability}

\begin{figure}
    \centering
    \includegraphics[width=\linewidth]{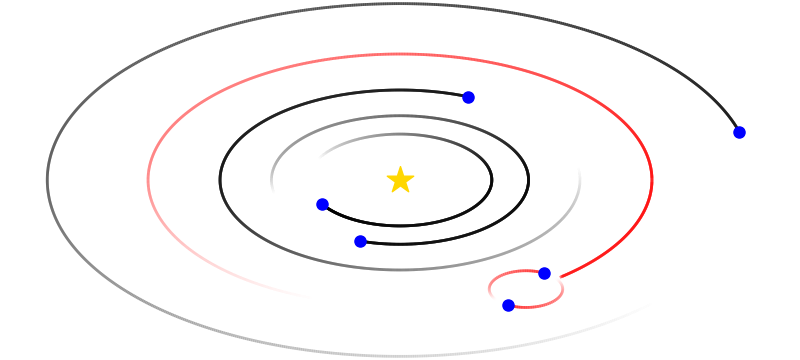}
    \caption{An illustrative cartoon showing a multiplanetary system with a dynamical gap viewed from an oblique angle, with a binary planet (red orbits) filling the dynamical gap.}
    \label{figure: diagram}
\end{figure}



\subsection{Target selection}
Our simulation targets were selected by taking the list of known exoplanetary systems from the composite data table published on the Exoplanet Archive as of June 5, 2024 \citep{Christiansen2025NASAExoplanetArchivea}.  We first filtered these systems to those we considered compact multiplanetary systems by requiring each system to have at least four transiting planets.  A total of 66 known planetary systems meet these criteria.

\begin{figure*}
    \centering
    \includegraphics[width=\linewidth]{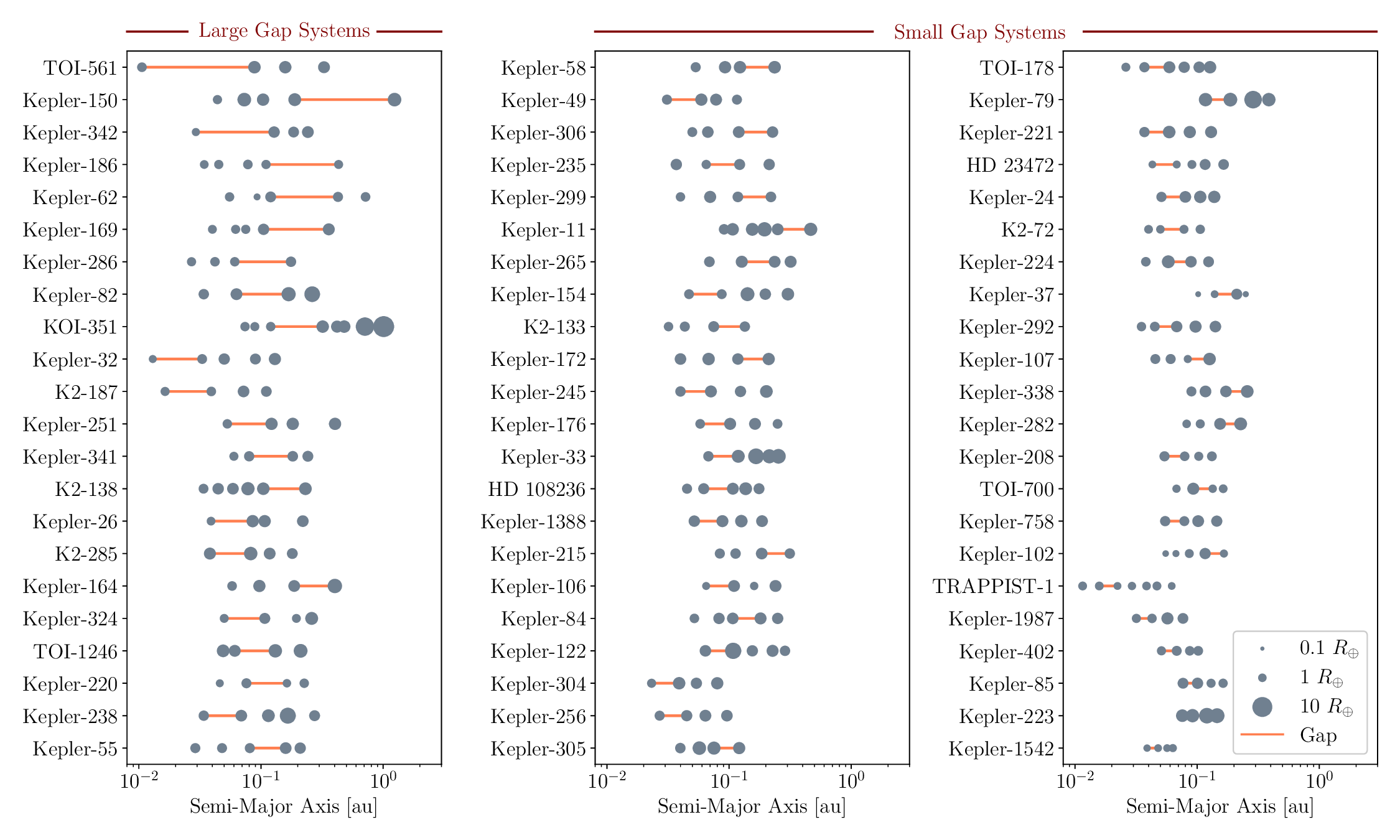}
    \caption{The 66 known planetary systems considered in this study.  The size of the dots corresponds to the radii of the planets, and the orange line identifies the largest gap in each system.  The systems are ranked by the size of the gap, which is defined as the ratio of the semi-major axes of the bounding planets.}
    \label{figure: known_sys}
\end{figure*}

Dynamical interactions are expected to evolve multiplanetary systems into an architecture where logarithms of orbital separations are uniformly distributed \citep{Mishra2021NewGenerationPlanetary}. We identified the largest gap between planets by finding which pair of adjacent planets has the largest ratio of semi-major axes.  While we simulate every system that satisfies our criteria regardless of the size of the dynamical gap, we separate our sample into systems with a ``large gap'' ($a_{i+1}/a_{i} > 1.95$, chosen to be the largest 1/3 of our sample) and those with a ``small gap.''  These 66 systems, along with their respective gaps and categorizations, are presented in Figure \ref{figure: known_sys}.

\subsection{Dynamical simulations}
For each target system, we run simulations of two different scenarios. We simulate both a ``single injection'' scenario, where we simulate the host star, all known planets, and an additional planet in the largest gap between known planets, and a ``binary injection'' scenario, where we instead inject a binary planet pair into the largest gap.  We performed Monte Carlo simulations with 3000 runs for each system under the single-injection scenario and 6000 runs for the binary-injection scenarios to determine which fraction of each scenario-system pair was stable.  The larger number of binary runs is to compensate for the lower survival rate of binary systems.  We initialize our N-body simulations of these systems by drawing parameters from the priors described in Section \ref{section: priors}.

Our code uses the \texttt{rebound} Python package \citep{Rein2012REBOUNDOpensourceMultipurpose} to perform the N-body simulation.  We used the \added{\texttt{trace} integrator \citep{Lu2024TraceCodeTimereversible}, a hybrid integrator.  \texttt{trace} was chosen because we desired a fast symplectic integrator but could not assume that a central, starward force dominated our systems with added binaries. \texttt{trace} treats close encounters, such as a tight binary, accurately while still using a fast Wisdom-Holman map on the remaining bodies. While the bodies in our system have a radius for the purposes of tracking collisions, the rest of the time} they are treated as point particles, notably meaning that we do not consider tidal dissipation due to close encounters. Each run was simulated for \added{$10^6$} years at a time step \added{set to 5\% of the period of the innermost planet of each system}.  We additionally implemented an early stopping mechanism to save computational time of clearly unstable runs by stopping any run for which a planet a) became gravitationally unbound, b) entered an orbit with a semi-major axis $a > 10$ au, or \added{c) collided with another planet or the star}.  Additionally, to ensure that the binary planet remains a bound pair, we stopped all binary injection scenarios where the distance between the injected planets exceeded half the original semi-major axis of the binary system with respect to the star.  Due to phase randomization over time, this criterion is practically guaranteed to be reached if the planets are no longer bound.

\subsection{Parameter space probed}
\label{section: priors}
When initializing the planetary systems, our default priors for the stellar mass, planetary masses, planetary semi-major axes, planetary radiii, and planetary eccentricities of the known planets in the system were to take a uniform distribution between the upper and lower uncertainty bounds reported in the Exoplanet Archive Composite Data Table for the system, if they existed.  A complete table of the references used in determining our priors is presented in Appendix \ref{datasources}.

These data were not available for all 66 exoplanetary systems, so we implemented fallback priors for those cases.  When the planetary mass is not known, \added{as is the case for 25 of our known planets}, we used mass-radius relations as a fallback prior.  We drew a random planetary radius from a uniform distribution between the known upper and lower uncertainty bounds in radius. Then, we used the \texttt{forecaster} \citep{Chen2016PROBABILISTICFORECASTINGMASSES} Python package to calculate the mass of the planet.  We use rejection sampling to ensure that any drawn masses are consistent with known upper mass bounds.  

Every planet in our sample has a known, most likely semi-major axis, but the uncertainty bounds on these estimates are missing for a few planets.  In these cases, \added{we assumed that the catalog value of the semi-major axis is an accurate reflection of the planetary period and rescaled the semi-major axis according to the stellar mass draw for the system}.  We also assumed an uncertainty of $\pm 20\%$ for planets without planetary mass bounds and the three systems without listed stellar mass bounds.
\added{For planets without known eccentricity we set the eccentricity to zero as detected planets without listed eccentricities are assumed to be disproportionately likely to have eccentricities that are unmeasurable and thus close to zero.}



To reduce the number of free parameters, we assume that the non-injected planets all orbit in a plane and thus have an inclination of 0. We also randomize the longitude of the ascending node and the argument of pericenter by drawing from a uniform distribution throughout the entire possible domain from $0$ to $2\pi$.

When injecting additional planets into the system, we have no prior knowledge of the planetary parameters. Therefore, we must make some assumptions when choosing our priors.  When we inject a single planet, we draw its mass from a uniform distribution from $0.38 \text{ M}_\oplus$ to $72 \text{ M}_\oplus$. These bounds were \added{originally chosen because they were near the masses of the least and most massive planets in our initial sample (a subset of the 66 systems), but with our current sample should be treated as arbitrary}. The logarithm of the semi-major axis was drawn from a uniform distribution bounded by the semi-major axes of the adjacent planets. The eccentricity of the injected planet was drawn from a uniform distribution between 0 and 1. Unlike for known planets, we \added{isotropically} draw the inclination of the injected \added{planet between $\cos i = -1$ and $\cos i = 1$}.

When injecting a binary planet into the system, we determine the masses of the injected planets by first drawing the total mass, $m_\text{tot}$, from a uniform distribution from $0.38 \text{ M}_\oplus$ to $72 \text{ M}_\oplus$.  We draw the mass fraction of the smaller planet in the pair, $q$, from a uniform distribution between 0 and 0.50.  Similarly to the single injection case, we draw the semi-major axis of the center of mass of the binary system around the star, $a_\text{COM}$, from a uniform distribution in logarithmic space, bounded by the semi-major axes of the adjacent planets.  We draw the eccentricity of the center of mass of the binary system around the star, $e_\text{COM}$,  from a uniform distribution between \added{0 and $e_{\max}$, where we set $e_{\max} = \min(a_\text{out}/a_\text{in} - 1, 1 - a_\text{in}/a_\text{out})$ to exclude orbits guaranteed to be crossing, with a stricter orbit crossing check done later}.  The eccentricity of the planets in orbit around each other, $e_\text{bin}$, is instead drawn uniformly between 0 and 1.  We then set the mass of each planet as follows: $m_1 = qm_\text{tot}$, $m_2 = (1-q)m_\text{tot}$.  We draw the initial semi-major axis between the two planets from a uniform distribution between 0 and the Hill radius of the binary system,
\begin{equation} \label{eqn: hill}
    r_\text{hill} = a_\text{COM} (1 - e_\text{COM}) \left(\frac{m_\text{tot}}{3M_\text{star}}\right)^{1/3}
\end{equation}
We draw the inclinations of the binary center of mass and the \added{mutual inclination from a uniform distribution between $\cos i = -1$ and $\cos i = 1$.}

Lastly, for all scenarios, we reject and resample any systems that are not stable for at least 5 years.  The reason for this is twofold.  Firstly, while we are initializing each of these systems using \added{Keplerian} elements and using positions and velocities derived from the solution to the two-body problem, these systems are significantly more complex than 2-body systems, especially for the binary injected systems where we assume the state vectors of the planets are equal to the sum of the two-body \added{Keplerian} orbit of the center of mass around the star and the two-body \added{Keplerian} orbit of the planets around each other. Not rejecting short-term unstable systems would lead to the misrepresentation of certain initial states as overly unstable, such as when the binary planets are initialized along a radial vector from the star, even though the true cause of the instability is the breakdown of the sum-of-Keplerian-orbits assumption.  Secondly, the scope of this paper is similarly limited to examining the long-term stability of these planetary systems, given that they are already in specific orbital configurations.  Investigations into how planets initially form in these orbital configurations are thus outside the scope of this paper. \added{We similarly reject and resample any systems initialized on crossing orbits (defined by the periastron and apastron of the center-of-mass of the injected planets and their neighbors, broadened by the initial binary separation in the case of binary injection runs).}




\section{Results} \label{sec: results}
\subsection{Population-Level Stability}
We consider a system to be dynamically stable if a) all known and injected planets remain gravitationally bound to the system over the entire simulation, b) if the system had an injected binary pair, that pair of planets remained gravitationally bound to each other, \added{and c) no collisions occurred within the system}. 
Criterion (b) is a stricter version of the corresponding early stopping criterion. We determine that planets are gravitationally unbound if, at any point in the simulation, they reach a separation greater than 110\% of their mutual hill radius.  While not in our final analysis, we also considered a stability factor that required all semi-major axes in the system to remain within a threshold of their original values.  This criterion was ultimately discarded as there was no natural choice for what value to pick for the threshold and the vast majority of our stable systems do not show any appreciable migration in semi-major axis.


\begin{figure*}
    \centering
    \includegraphics[width=0.7\linewidth]{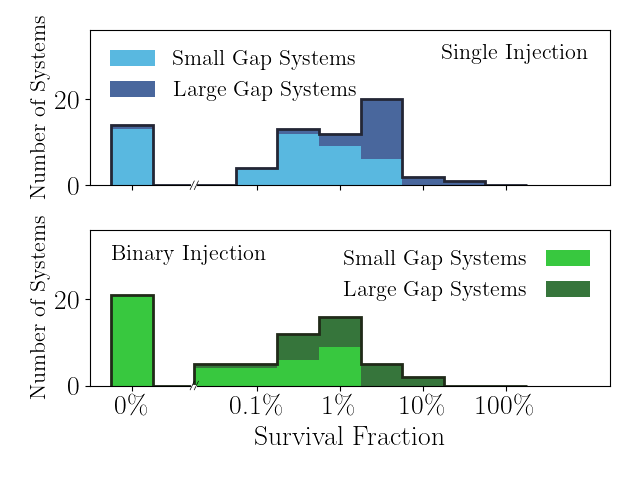}
    \caption{Histogram of fraction of stable systems single injection (top), and binary injection (bottom).  The darker regions of the histograms indicate systems with a large dynamical gap.  Overall, the distributions of single- and binary-injection stability are similar. }
    \label{figure: hist}
\end{figure*}

\begin{figure*}
    \centering
    \includegraphics[width=0.7\linewidth]{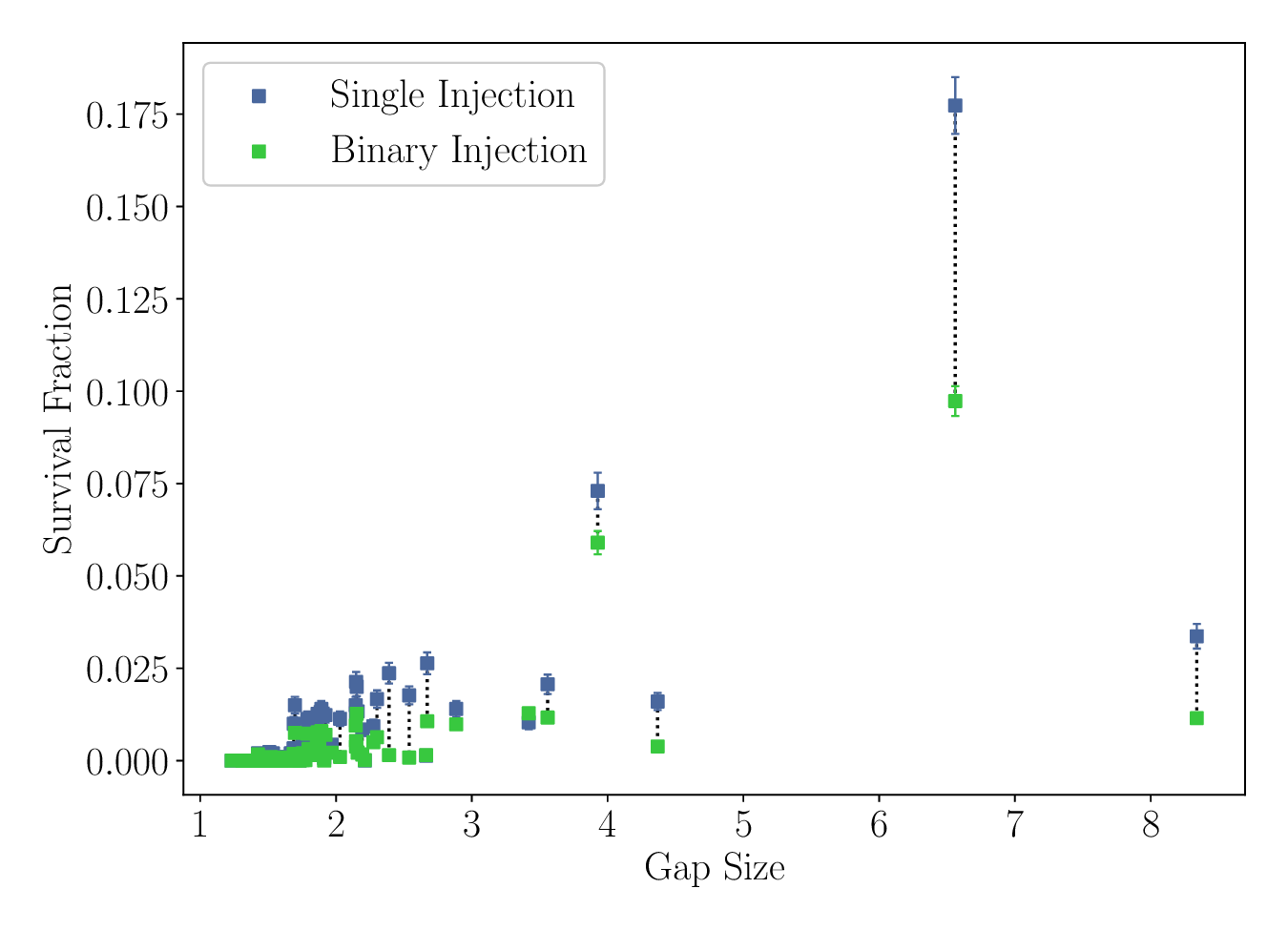}
    \caption{Survival fraction for both single injection (blue) and binary injection (orange) as a function of gap size.  Gap size is defined as the ratio between the semi-major axis of the outer planet bounding the gap and the semi-major axis of the inner planet bounding the gap.  The single- and binary-injection cases for the same system are connected by a dotted line.  Error bars are $\pm 1\sigma$.   As the gap size increases, the survival fraction also tends to increase.}
    \label{figure: scatter}
\end{figure*}

We present a summary histogram in Figure~\ref{figure: hist} that shows the fraction of simulations that were stable for each system.  Overall, although our simulations with a single injected planet were generally more stable, the difference was slight.  We also note that, as expected, our most stable systems with injected planets are predominantly those with a large gap.  This is an intuitive result, as these injected planets will be subject to relatively smaller perturbations, \added{and in turn purturb the known system less}, than in systems with a smaller gap.  This trend is further elucidated by Figure~\ref{figure: scatter}, which shows that there is a correlation between the gap size (as defined by the ratio of the semi-major axis of the planets bounding the gap) and both the binary and single injection survival rates.

\begin{figure*}
    \centering
    \includegraphics[width=0.8\linewidth]{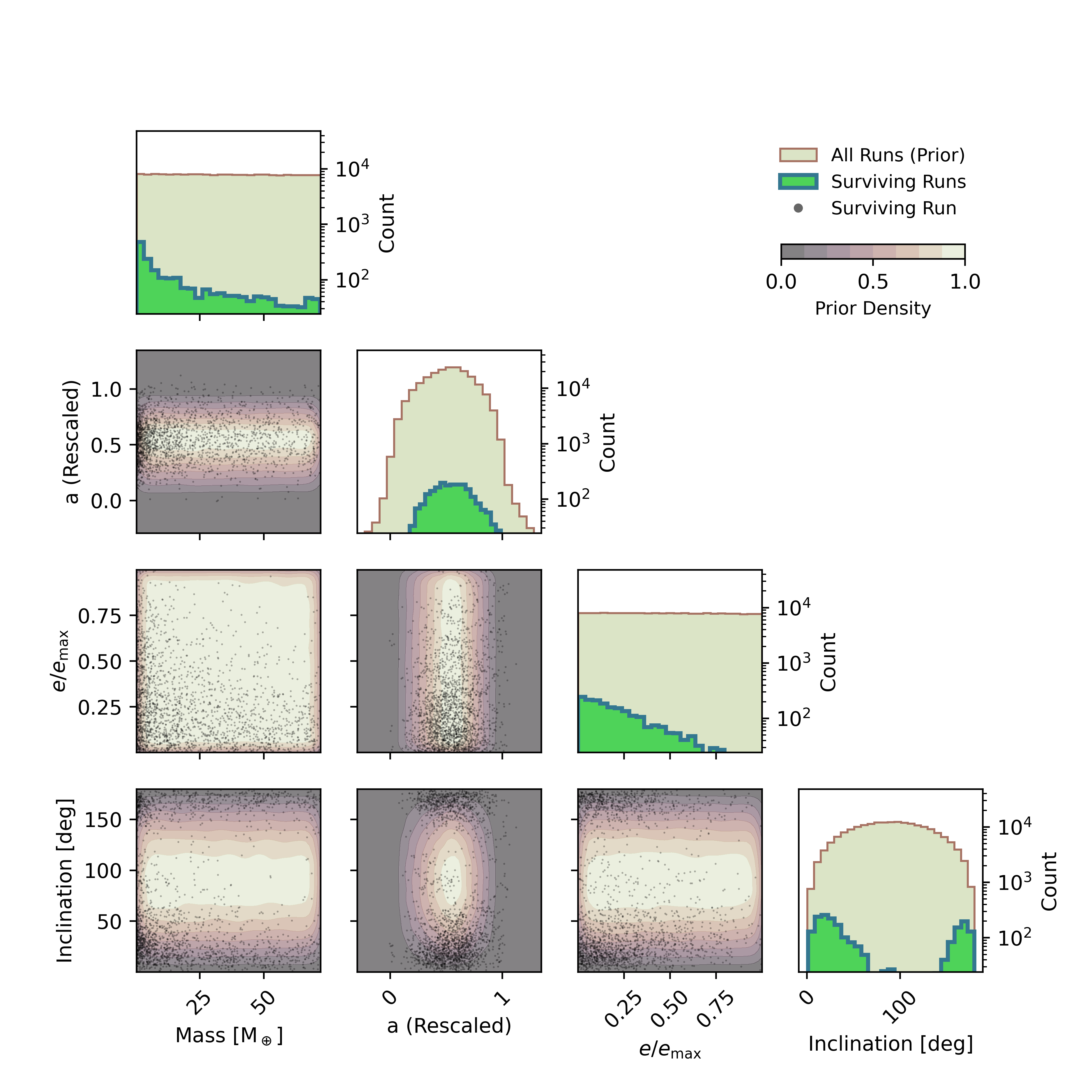}
    \caption{Corner plot showing the prior (tan) and surviving (gray points, green histogram) distribution of single planets injected into dynamical gaps, marginalized over all systems.  The semi-major axis has been normalized such that 0 corresponds to the inner boundary of the dynamical gap, 1 to the outer boundary, and linear changes in the normalized parameters correspond to logarithmic changes in semi-major axis. \added{The eccentricity has been normalized by $e_\text{max}$, the maximum eccentricity at the drawn semi-major axis which does not result in crossing orbits.}}
    \label{figure: corner_sin}
\end{figure*}
\added{For each unstable system we track the reason we mark the system as unstable.  Across all systems, collisions are the dominant marker of instability--accounting for $\sim$$99\%$ of runs.  Planet scattering causing planets to become unbound or on orbits larger than 10 au occurs in $\sim$$0.3\%$ of runs, while the remaining $\sim$$1\%$ of runs are stable.  The collisions can be further broken down by which bodies collide---the injected planet hits a known planet in $\sim$$43\%$ of runs, a known planet hits the star in $\sim$$32\%$ of runs, two known planets collide in $\sim$$21\%$ of runs, and the injected planet hits the star in $\sim$$2.5\%$ of runs.} 

We present a corner plot showing the initial orbital parameters of the singly injected planet marginalized over all systems in Figure \ref{figure: corner_sin}.  There are \added{2114} stable configurations and \added{195886} unstable configurations.  Note that the semi-major axis has been rescaled so that we can marginalize over the systems---a rescaled semi-major axis of 0 or 1 corresponds to an orbit at the inner or outer edge of the gap, respectively, with interpolation between these end-states done logarithmically.  We similarly rescale the eccentricity to normalize by the maximum eccentricity that does not result in a crossing orbit given the drawn semi-major axis.  Our simulations moderately favor planets with lower masses and planets with orbits spaced far from their nearest neighbour.  Additionally, our simulations strongly favor planets with inclinations of near 0 (coplanar prograde) and $\pi$ (coplanar retrograde).

These results are mostly intuitive. Lower mass planets are less likely to disturb the other planets in the system---\added{an explanation corroborated by the fact that the number of known planets colliding with other known planets or the star increases with the injected planet mass in a manner that mostly explains the difference in survival rates between low mass and high mass injections.}  \added{Eccentricities near 0 are strongly preferred to eccentricities close to the maximum obtainable without causing initialized orbits to cross.  The share of outcomes is largely the same between different eccentricities.  We note that this effect largely disappears when we do not rescale eccentricity, suggesting the eccentricity relative to the architecture of the system is far more important than the eccentricity alone}

Additionally, keeping inclinations near 0 (or $\pi$) keeps the system dynamically cool, which intuitively will increase the survival rate.  \added{The removal of planets injected on non-co-planar orbits is suggestive of a quadrupole Lidov-Kozai effect being responsible.  This story, however, is complicated by a) that these planetary systems are compact and not strongly hierarchical, b) that many of the planets evolve onto crossing orbits faster than the quadrupole Lidov-Kozai timescale, and c) that eccentricity and inclination in the removed planets do not trade off in a manner that conserves $\sqrt{1-e^2}\cos i$ \citep{Lidov1962EvolutionOrbitsArtificial, Kozai1962SecularPerturbationsAsteroids, Naoz2016EccentricKozaiLidovEffect}.  The octupole Lidov-Kozai effect, which can have notable effects even in near-coplanar orbits, may play a role in some of our systems, but is suppressed by construction in $\sim$$80\%$ of them.  The strength of the octupole term, $\epsilon = ({a_\text{inj}}/{a_\text{out}})(e_\text{out} / (1 - e_\text{out}^2))$ \citep{Li2014CHAOSTESTPARTICLE}, is zero for the systems for which we did not have eccentricity constraints for the known planets.  For the systems with known eccentricity constraints, the octupole strength is non-negligible ($\epsilon > 0.001$), but there is no consistent correlation within a single system between the octupole strength of the random draw of orbital parameters and the survival rate of that draw.  We also note that there is an asymmetry between the prograde and retrograde orbits.  While both prograde and retrograde orbits have similar survival rates very near coplanar orbits, prograde orbits are more stable in the intermediate ($\sim$20$^\circ$ - $\sim$70$^\circ$) regime.  Retrograde orbits cause more collisions between known planets and the star, and collisions between the injected planet and the star occur predominantly on retrograde orbits.}


To a large extent, the semi-major axis trend is intuitive, with the most stable systems having the injected planet located as far away from the other planets as possible. At first glance, however, the number of survivors near the edge of the dynamical gap is surprising. \added{This phenomenon is partly caused by the fact that in systems with large enough gaps, planets can be close to the edge of normalized injection range without being physically close to other planetary orbits, in terms of mutual hill radii.  Survival rates generally decrease as the hill-radii separation between the injected and bounding orbits decreases; however, they spike again below one hill radius.  This is caused by the injected planets being placed in stable co-orbital configurations with their neighbors---roughly half of these co-orbiting planets are in tadpole orbits and the other half are in horseshoe orbits.  The semi-major axis and the eccentricity are not independent in how they are constrained by initial orbit-crossing test---orbits with higher eccentricity cross the orbit of the neighboring planets unless the planet is located towards the center of the gap.}

\begin{figure*}
    \centering
    \includegraphics[width=0.97\linewidth]{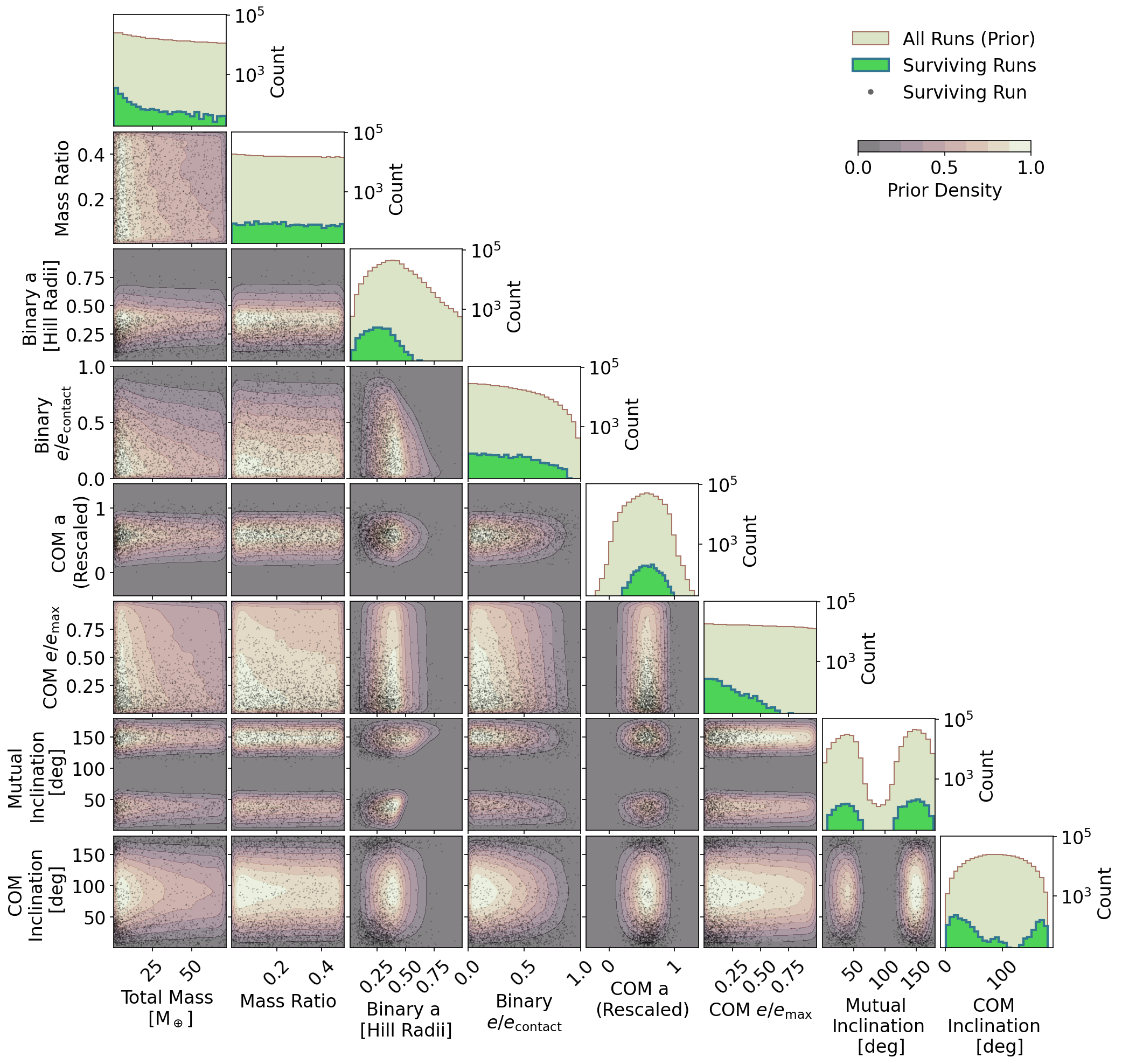}
    \caption{Corner plot showing the prior (tan) and surviving (gray points, green histogram) distribution of binary planets injected into dynamical gaps, marginalized over all systems.  The semi-major axis has been normalized such that 0 corresponds to the inner boundary of the dynamical gap, 1 to the outer boundary, and linear changes in the normalized parameters correspond to logarithmic changes in semi-major axis.  The separation between planets is in units of hill radii (See Eqn \ref{eqn: hill}). \added{The eccentricity has been normalized by $e_\text{max}$, the maximum eccentricity at the drawn semi-major axis which does not result in crossing orbits.  Similarly, the binary eccentricity has been normalized by $e_\text{contact}$, the maximum eccentricity at the drawn $a_\text{bin}$ that does not cause the planets to collide on its initial orbit.}}
    \label{figure: corner_bin}
\end{figure*} 
We additionally present a corner plot of the initial orbital parameters marginalized over all systems we simulated under binary injection in Figure \ref{figure: corner_bin}.  The semi-major axis variable is rescaled in the same manner as in the single injection case, and the initial semi-major axis of the binary system is in units of mutual hill radii.  Marginalization over the parameters was done using these rescaled binary separation and semi-major axis variables.   There are \added{2,011} stable configurations and \added{393,989} unstable configurations in our simulations. \added{Similar to the single injection case, we track why each system is determined to be unstable.  Collisions again account for the vast majority ($\sim$92\%) of the cases, with the binary becoming separated in $\sim$7.5\% of cases, and a planet being scattered beyond a semi-major axis of 10 au in $\sim$0.1\% of cases, with survival occurring in the remaining $\sim$0.5\% of cases.  Breaking down the collisions, $\sim$$50\%$ of runs end with the injected planets colliding with each other, $\sim$19\% of runs end with one of the injected planets hitting a known planet, $\sim$12\% end with a known planet hitting the star, and $\sim$10\% end with two known planets colliding with each other.  Compared with the single injection case, the ratios among the known planet-star, known-known planet, and injected-known planet outcomes is highly similar, with a small shift towards more injected planet-known planet collisions and away from known planet-star collisions---a trend that holds at a system-by-system level of analysis.}

The center-of-mass parameters, which are largely analogous to the single-injection case, have simular constraints to those in the single-injection case.  We find that there is again a moderate preference for planets to be located near the center of the gap, \added{with the number of injected-known planet collisions increasing close to the edge of the gap, again with a very small number of binary planets co-orbiting with their neighbors.   Similarly, there is again a moderate preference for lower mass injections, with higher total mass between the injected planets causing an increase in the number of known planet-star collisions.  The center-of-mass eccentricity is again constrained to prefer eccentricities which are small relative to the maximum eccentricity that does not result in crossing orbits.  The inclination of the center-of-mass also follows the trend of the single injection inclination constraints, with co-planar orbits being preferred, and with intermediate inclination prograde orbits being preferred to intermediate inclination retrograde orbits, largely driven by polar and intermediate inclination retrograde orbits causing a disproportionate number of known planet-star collisions and collisions between the injected pair.}



\added{The strongest constraint on the binary orbital parameters, in contrast to the single injection orbital parameters, is on the mutual inclination of the binary orbit.  While a lot of the constraints are set by the 5 year startup excluding much of the region between $\sim$60 and $\sim$120 degrees, the long simulations further exclude near-polar orbits beyond this prior as the number of collisions between the injected planets spikes in this region.  This 5 year startup also favors retrograde over prograde orbits by a factor of $\sim$2, an effect that might be caused by the shorter timeframe the injected planets remain in close alignment with the other planets in a retrograde orbit \citep{Gayon2007RetrogradeResonancesCompact, Gayon2008AreRetrogradeResonances}.  These trends in inclination are evocative of the observed trends in the orbits of the irregular satellites of the Solar System giant planets where inclinations between $\sim$50 and $\sim$140 degrees are largely unstable due to Lidov-Kozai effects induced by stellar perturbations \citep{Nesvorny2003OrbitalCollisionalEvolution}.  Unlike the single injection case, the binary injected planets are in a hierarchical system and thus more clearly subject to quadrupole Lidov-Kozai effects.  We note that the timescale of these effects $\tau \sim (P_\text{COM}^2/P_\text{bin} )((M_* + m_\text{tot})/M_*)(1-e_\text{COM}^2)^{3/2}$ is short relative to our 5 year start-up time so most systems disrupted by this process are rejected and resampled by the start up window.}

\added{The semi-major axis of the binary shows a moderate constraint for tighter orbits.  In general, binary orbits which are wider are more likely to become unbound with the injected planets separating or colliding with each other as interactions with the star and other planets push the eccentricity of the binary up.  When we rescale the binary eccentricity by $e_\text{contact} = 1 - (R_1 + R_2)/a_\text{bin}$, the maximum eccentricity that does not place the binary pericenter inside the injected planets, leading to a collision, we find that the allowed eccentricity is mostly flat (within a factor of 2) with respect to the start-up prior, which preferred smaller eccentricities.  When we control for the binary semi-major axis to account for the fact that the only systems that survive the 5 year start-up with eccentricities near this contact limit are ones in tight, low $a_\text{bin}$, orbits the factor of 2 disappears.}
\added{There appear to be no constraints on the mass ratio between the injected planets indicating that the exomoon and binary planet cases are largely analogous.  There similarly is no difference in collision outcomes between the different mass ratios.}


\subsection{Individual System Stability}
\begin{figure*}
    \centering
    \includegraphics[width=0.7\linewidth]{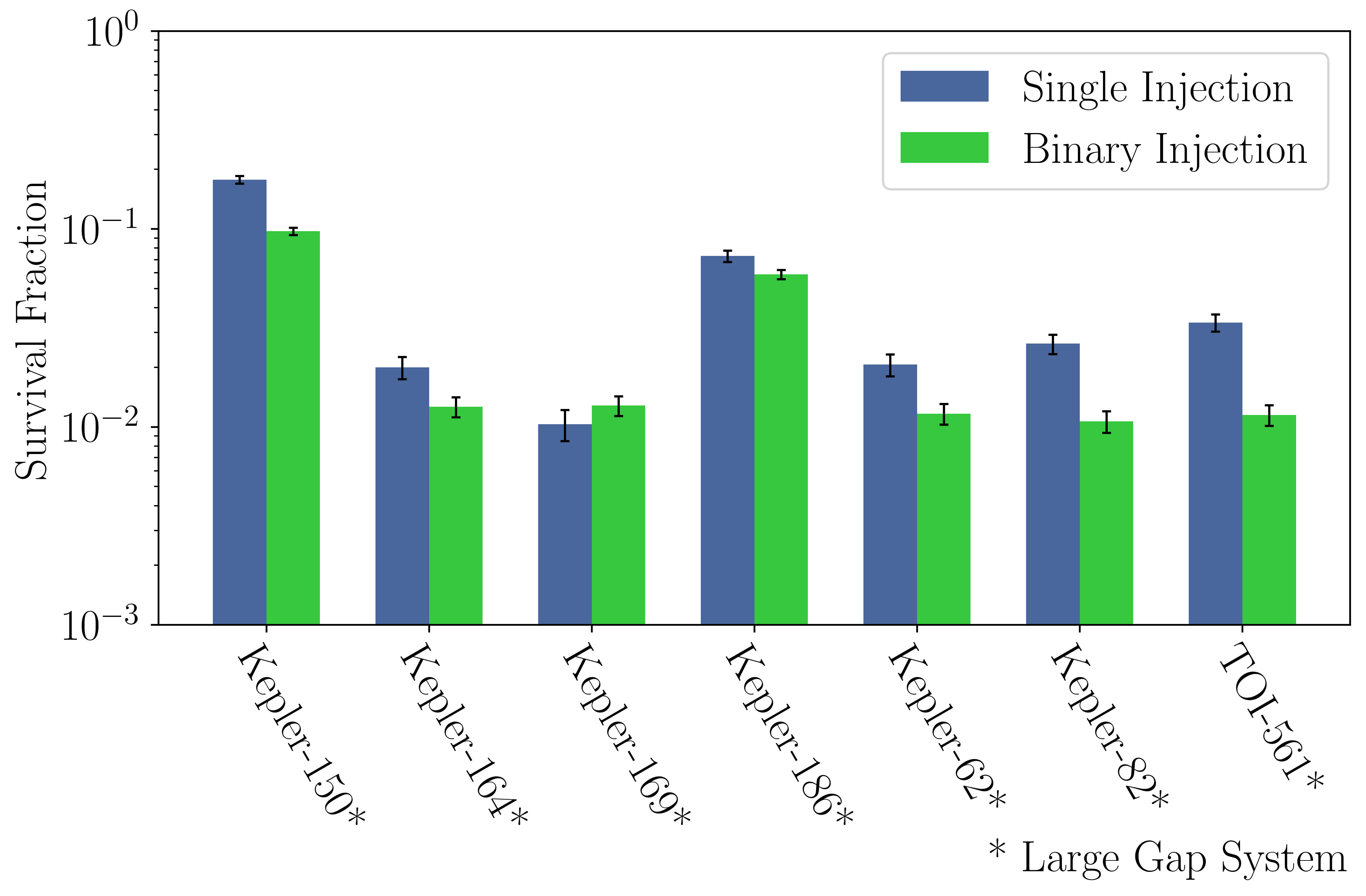}
    \caption{Fraction of stable systems for those systems which had a binary injection stability that exceeded 1\%.  Error bars are $\pm1\sigma$, and an asterisk next to the system name indicates the system is a large gap system (gap size in the top 1/3 of the sample).}
    \label{figure: survival_frac}
\end{figure*}

\added{Of the systems simulated, seven (Kepler-150, Kepler-164, Kepler-169, Kepler-186, Kepler-62, Kepler-82, TOI-561) have a binary injection survival rate of at least 1\%.  Notably all, of these systems are large gap systems, meaning they have dynamical gap sizes in the upper third of our sample.  Two systems in particular stand out. Kepler-150 has the highest binary injection survival rate of any system at $9.73\% \pm 0.40\%$.  Kepler-169 also stands out as the only system with a non-negligible binary injection survival rate ($1.28\% \pm 0.19\%$) higher than that of the single injection survival rate ($1.03\% \pm 0.15\%$).  We note that these individual system level results are more sensitive to modeling assumptions than the population level results, which in contrast to the individual system level results, have remained stable as we refined our model.}

\section{Discussion and Conclusion}
\label{sec: conclude}
We have used N-body simulations to model the dynamical stability of potential as-yet-undiscovered single and binary planets that might exist in the dynamical gaps of known compact multiplanetary systems.  We find a small but well-defined region of parameter space where injected planets can be dynamically stable.  Notably, we find that while single-injected planets are overall more stable than binary-injected planets, the difference in stability is limited to a factor of two.  

We overall find reasonably strong, intuitive, and similar constraints on the injected single planets and the center-of-mass motion of the injected binaries, \added{with these two cases being largely analogous}.  Low mass, low eccentricity, and coplanar inclinations are strongly preferred, with a preference for inhabiting the center of the dynamical gap.  \added{We additionally find that in the injected binary system, stability prefers the binary to have a low mutual inclination and a low semi-major axis.  The eccentricity of the binary orbit is, in contrast, poorly constrained and the mass ratio is similarly flat---indicating little to no difference dynamically between the exo-moon and binary planet regime.}

We note that, in this study, we ignored the tidal effects on individual planets.  These tidal forces would dissipate energy, likely making the binary planets more tightly bound \citep{Ochiai2014EXTRASOLARBINARYPLANETS}, \added{potentially reducing the number of separations and increasing collisions.}  Similarly, all planets but the injected one(s) had an inclination of 0.  If we were to relax this assumption and introduce an angular-momentum deficit, we might see a relaxation of the tight inclination constraints on the injected planet.  We additionally do not consider the relative formation efficiencies between single and binary planets.  \added{While, as demonstrated by Pluto-Charon, we know that binary planet formation, at least for dwarf planets, is possible,}  we do not know how frequently such pairs actually form, especially in the compact environments we simulate here.  \added{Some rapid timescale effects hidden by our 5 year start-up window, such as the short timescale of quadrupole Lidov-Kozai, may indicate that formation of binary planets in these compact systems is difficult.}  We only note that should such a binary have formed in the dynamical gaps considered in this study, its odds of survival are similar to those of an additional single planet.

We also note that most of the stable injected planets have low to no inclination and are thus likely to transit.  While there has been some work modeling the shapes of binary-planet transits \citep{Chakraborty2022TransitDurationTiming}, this modeling was done only at small orbital separations. In contrast, we find stable orbital configurations at far wider separations.  Expanding such modeling, combined with the knowledge that most stable binary planets would be orbitally aligned with their systems, can motivate archival surveys designed to constrain the occurrence rates of binary planets in these compact multi-planetary systems.

\label{section:DiscussionAndConclusion}

The code used to perform these simulations, the simulation data which were created, and the code used to create the figures analyzing the simulation data for this paper are available on Zenodo\footnote{\href{https://doi.org/10.5281/zenodo.23697083}{https://doi.org/10.5281/zenodo.23697083}} \citep{Cukier2026CodeSimulationOutput}.

\section{Acknowledgements}

T.D. acknowledges support from the McDonnell Center for the Space Sciences at Washington University in St. Louis.

The authors acknowledge the Research Infrastructure Services (RIS)\footnote{https://ris.wustl.edu/} at Washington University in St. Louis for providing computational resources and services
needed to generate the research results delivered in this paper. This research was done using services provided by the OSG Consortium, which is supported by the National Science Foundation awards \#2030508 and \#2323298.  This work used the Open Storage Network and the Open Science Pool at OSG through allocation PHY260390 from the Advanced Cyberinfrastructure Coordination Ecosystem: Services \& Support (ACCESS) program, which is supported by U.S. National Science Foundation grants \#2138259, \#2138286, \#2138307, \#2137603, and \#2138296.  This research has made use of the NASA Exoplanet Archive, which is operated by the California Institute of Technology, under contract with the National Aeronautics and Space Administration under the Exoplanet Exploration Program.  LLM tools (Opus 5, 5.5, Fable 5, 5.1) were used to refine our simulation code, find references, manage simulation run logistics, proofread the manuscript, and help create data visualizations. 


\begin{contribution}
WC led the analysis, interpretation, and writing for this project. TD conceived of this project and supervised and contributed to the analysis, interpretation, and writing for this project.
\end{contribution}

\software{\texttt{numpy} \citep{Harris2020ArrayProgrammingNumPy}, \texttt{matplotlib} \citep{Hunter2007Matplotlib2DGraphics}, \texttt{scipy} \citep{Virtanen2020SciPy10Fundamental}, \texttt{rebound} \citep{Rein2012REBOUNDOpensourceMultipurpose}, \texttt{forecaster} \citep{Chen2016PROBABILISTICFORECASTINGMASSES}, Open Science Grid \citep{osg07, osg09, osg06, osg15}, ACCESS \citep{Boerner2023ACCESSAdvancingInnovation}}
\appendix
\section{Data Sources}
\label{datasources}

\begin{longrotatetable}

\begin{deluxetable*}{lcccccc}
\tablecaption{Data sources for orbital parameters of known planets.}

\tabletypesize{\tiny}
\tablehead{
\colhead{System} & \colhead{Stellar Mass} & \colhead{Mass} & \colhead{Radius} & \colhead{e} & \colhead{a} \\
}
\startdata
HD 108236 & \shortstack{\citet{Bonfanti2021CHEOPSObservationsHD}} & \shortstack{Exoplanet Archive Calculated} & \shortstack{\citet{Bonfanti2021CHEOPSObservationsHD}} & \shortstack{\citet{Bonfanti2021CHEOPSObservationsHD}} & \shortstack{\citet{Bonfanti2021CHEOPSObservationsHD}}  \\
HD 158259 & \shortstack{\citet{Hara2020SOPHIESearchNorthern}} & \shortstack{\citet{Hara2020SOPHIESearchNorthern}} & \shortstack{EXOFOP TESS TOI} & \shortstack{N/A} & \shortstack{N/A} \\
HD 191939 & \shortstack{\citet{Lubin2022TESSKeckSurveyIX}} & \shortstack{\citet{Orell-Miquel2023HD191939Revisited}} & \shortstack{\citet{Orell-Miquel2023HD191939Revisited}} & \shortstack{\citet{Orell-Miquel2023HD191939Revisited}} & \shortstack{\citet{Orell-Miquel2023HD191939Revisited}}  \\
HD 219134 & \shortstack{\citet{Gillon2017TwoMassiveRocky}} & \shortstack{\citet{Gillon2017TwoMassiveRocky}} & \shortstack{\citet{Gillon2017TwoMassiveRocky}} & \shortstack{\citet{Gillon2017TwoMassiveRocky}} & \shortstack{\citet{Gillon2017TwoMassiveRocky}} \\
HD 23472 & \shortstack{\citet{Barros2022HD23472Multiplanetary}} & \shortstack{\citet{Barros2022HD23472Multiplanetary}} & \shortstack{\citet{Barros2022HD23472Multiplanetary}} & \shortstack{\citet{Barros2022HD23472Multiplanetary}} & \shortstack{\citet{Barros2022HD23472Multiplanetary}}  \\
HD 3167 & \shortstack{\citet{Bonomo2023ColdJupitersImproved}} & \shortstack{\citet{Bonomo2023ColdJupitersImproved}} & \shortstack{\citet{Bonomo2023ColdJupitersImproved}} & \shortstack{\citet{Bonomo2023ColdJupitersImproved}} & \shortstack{\citet{Bonomo2023ColdJupitersImproved}} \\
HIP 41378 & \shortstack{\citet{Berardo2019RevisitingHIP41378}} & \shortstack{Exoplanet Archive Calculated} & \shortstack{EXOFOP TESS TOI \\ \citet{Vanderburg2016FivePlanetsTransiting}} & \shortstack{\citet{Grouffal2022RossiterMcLaughlinDetection9month}} & \shortstack{N/A}  \\
K2-133 & \shortstack{\citet{Wells2019ValidationTemperateFourth}} & \shortstack{Exoplanet Archive Calculated} & \shortstack{\citet{Wells2019ValidationTemperateFourth}} & \shortstack{N/A} & \shortstack{\citet{Wells2019ValidationTemperateFourth}} \\
K2-138 & \shortstack{\citet{Lopez2019ExoplanetCharacterisationLongest}} & \shortstack{\citet{Lopez2019ExoplanetCharacterisationLongest}} & \shortstack{\citet{Lopez2019ExoplanetCharacterisationLongest}} & \shortstack{\citet{Lopez2019ExoplanetCharacterisationLongest}} & \shortstack{\citet{Lopez2019ExoplanetCharacterisationLongest}} \\
K2-187 & \shortstack{\citet{Mayo2018275Candidates149}} & \shortstack{Exoplanet Archive Calculated} & \shortstack{\citet{Adams2021UltrashortperiodPlanetsK2}} & \shortstack{N/A} & \shortstack{\citet{Livingston2018SixtyValidatedPlanets}}\\
K2-268 & \shortstack{\citet{deLeon202137NewValidated}} & \shortstack{Exoplanet Archive Calculated} & \shortstack{\citet{deLeon202137NewValidated}} & \shortstack{N/A} & \shortstack{\citet{Livingston2018SixtyValidatedPlanets} \\ N/A}  \\
K2-285 & \shortstack{\citet{Palle2019DetectionDopplerMonitoring}} & \shortstack{\citet{Palle2019DetectionDopplerMonitoring}} & \shortstack{\citet{Palle2019DetectionDopplerMonitoring}} & \shortstack{\citet{Palle2019DetectionDopplerMonitoring}} & \shortstack{\citet{Palle2019DetectionDopplerMonitoring}}  \\
K2-384 & \shortstack{\citet{Christiansen2022ScalingK2Statistical}} & \shortstack{Exoplanet Archive Calculated} & \shortstack{\citet{Christiansen2022ScalingK2Statistical}} & \shortstack{N/A} & \shortstack{N/A} \\
K2-72 & \shortstack{\citet{Dressing2017CharacterizingK2Candidate}} & \shortstack{Exoplanet Archive Calculated} & \shortstack{\citet{Dressing2017CharacterizingK2Candidate}} & \shortstack{\citet{Dressing2017CharacterizingK2Candidate}} & \shortstack{\citet{Dressing2017CharacterizingK2Candidate}}  \\
KOI-351 & \shortstack{\citet{Cabrera2014PlanetarySystemKIC}} & \shortstack{Exoplanet Archive Calculated \\ \citet{Santerne2016SOPHIEVelocimetryKepler}} & \shortstack{\citet{Cabrera2014PlanetarySystemKIC} \\ \citet{Shallue2018IdentifyingExoplanetsDeep}} & \shortstack{\citet{Weiss2024KeplerGiantPlanet}} & \shortstack{\citet{Cabrera2014PlanetarySystemKIC} \\ \citet{Weiss2024KeplerGiantPlanet}}   \\
Kepler-100 & \shortstack{\citet{Marcy2014MassesRadiiOrbits}} & \shortstack{\citet{Marcy2014MassesRadiiOrbits}} & \shortstack{\citet{Marcy2014MassesRadiiOrbits}} & \shortstack{\citet{VanEylen2015EccentricityTransitPhotometry}} & \shortstack{\citet{Weiss2024KeplerGiantPlanet}}  \\
Kepler-102 & \shortstack{\citet{Bonomo2023ColdJupitersImproved}} & \shortstack{\citet{Bonomo2023ColdJupitersImproved}} & \shortstack{\citet{Bonomo2023ColdJupitersImproved}} & \shortstack{\citet{Bonomo2023ColdJupitersImproved}} & \shortstack{\citet{Bonomo2023ColdJupitersImproved}} \\
Kepler-106 & \shortstack{\citet{Marcy2014MassesRadiiOrbits}} & \shortstack{\citet{Marcy2014MassesRadiiOrbits}} & \shortstack{\citet{Marcy2014MassesRadiiOrbits}} & \shortstack{\citet{Weiss2024KeplerGiantPlanet}} & \shortstack{\citet{Weiss2024KeplerGiantPlanet}} \\
Kepler-107 & \shortstack{\citet{Bonomo2023ColdJupitersImproved}} & \shortstack{\citet{Bonomo2023ColdJupitersImproved}} & \shortstack{\citet{Bonomo2023ColdJupitersImproved}} & \shortstack{\citet{Bonomo2023ColdJupitersImproved}} & \shortstack{\citet{Bonomo2023ColdJupitersImproved}}  \\
Kepler-11 & \shortstack{\citet{Lissauer2013AllSixPlanets}} & \shortstack{\citet{Lissauer2013AllSixPlanets}} & \shortstack{\citet{Lissauer2013AllSixPlanets}} & \shortstack{\citet{Lissauer2013AllSixPlanets}} & \shortstack{\citet{Lissauer2013AllSixPlanets}} \\
Kepler-122 & \shortstack{\citet{Morton2016FalsePositiveProbabilities}} & \shortstack{Exoplanet Archive Calculated \\ \citet{Hadden2014DensitiesEccentricities139}} & \shortstack{\citet{Hadden2014DensitiesEccentricities139} \\ \citet{Rowe2014ValidationKeplersMultiple}} & \shortstack{Q1-17 KOI Table} & \shortstack{Q1-17 KOI Table \\ \citet{Rowe2014ValidationKeplersMultiple}}  \\
Kepler-138 & \shortstack{\citet{Piaulet2023EvidenceVolatilerichComposition}} & \shortstack{\citet{Piaulet2023EvidenceVolatilerichComposition}} & \shortstack{\citet{Piaulet2023EvidenceVolatilerichComposition}} & \shortstack{\citet{Piaulet2023EvidenceVolatilerichComposition}} & \shortstack{\citet{Piaulet2023EvidenceVolatilerichComposition}}  \\
Kepler-1388 & \shortstack{\citet{Morton2016FalsePositiveProbabilities}} & \shortstack{Exoplanet Archive Calculated} & \shortstack{\citet{Morton2016FalsePositiveProbabilities}} & \shortstack{Q1-17 KOI Table} & \shortstack{Q1-17 KOI Table}  \\
Kepler-139 & \shortstack{\citet{Weiss2024KeplerGiantPlanet}} & \shortstack{\citet{Weiss2024KeplerGiantPlanet}} & \shortstack{\citet{Rowe2014ValidationKeplersMultiple} \\ \citet{Weiss2024KeplerGiantPlanet}} & \shortstack{\citet{Weiss2024KeplerGiantPlanet}} & \shortstack{\citet{Rowe2014ValidationKeplersMultiple} \\ \citet{Weiss2024KeplerGiantPlanet}} \\
Kepler-150 & \shortstack{\citet{Morton2016FalsePositiveProbabilities}} & \shortstack{Exoplanet Archive Calculated} & \shortstack{\citet{Rowe2014ValidationKeplersMultiple} \\ \citet{Schmitt2017SearchLostPlanets}} & \shortstack{N/A \\ Q1-17 KOI Table} & \shortstack{\citet{Rowe2014ValidationKeplersMultiple} \\ \citet{Schmitt2017SearchLostPlanets}} \\
Kepler-154 & \shortstack{\citet{Morton2016FalsePositiveProbabilities}} & \shortstack{Exoplanet Archive Calculated} & \shortstack{\citet{Morton2016FalsePositiveProbabilities} \\ \citet{Rowe2014ValidationKeplersMultiple}} & \shortstack{Q1-17 KOI Table} & \shortstack{Q1-17 KOI Table \\ \citet{Rowe2014ValidationKeplersMultiple}} \\
Kepler-1542 & \shortstack{\citet{Morton2016FalsePositiveProbabilities}} & \shortstack{Exoplanet Archive Calculated} & \shortstack{\citet{Morton2016FalsePositiveProbabilities}} & \shortstack{Q1-17 KOI Table} & \shortstack{Q1-17 KOI Table} \\
Kepler-164 & \shortstack{\citet{Morton2016FalsePositiveProbabilities}} & \shortstack{Exoplanet Archive Calculated} & \shortstack{Q1-17 KOI Table \\ \citet{Rowe2014ValidationKeplersMultiple}} & \shortstack{Q1-17 KOI Table} & \shortstack{Q1-17 KOI Table \\ \citet{Rowe2014ValidationKeplersMultiple}}  \\
Kepler-169 & \shortstack{\citet{Morton2016FalsePositiveProbabilities}} & \shortstack{Exoplanet Archive Calculated} & \shortstack{\citet{Rowe2014ValidationKeplersMultiple}} & \shortstack{Q1-17 KOI Table} & \shortstack{\citet{Rowe2014ValidationKeplersMultiple}}  \\
Kepler-172 & \shortstack{\citet{Morton2016FalsePositiveProbabilities}} & \shortstack{Exoplanet Archive Calculated} & \shortstack{\citet{Rowe2014ValidationKeplersMultiple}} & \shortstack{Q1-17 KOI Table} & \shortstack{\citet{Rowe2014ValidationKeplersMultiple}}  \\
Kepler-176 & \shortstack{Q1-17 KOI Table} & \shortstack{Exoplanet Archive Calculated \\ \citet{Hadden2014DensitiesEccentricities139}} & \shortstack{\citet{Morton2016FalsePositiveProbabilities} \\ \citet{Rowe2014ValidationKeplersMultiple}} & \shortstack{Q1-17 KOI Table} & \shortstack{Q1-17 KOI Table \\ \citet{Rowe2014ValidationKeplersMultiple}} \\
Kepler-186 & \shortstack{\citet{Quintana2014EarthSizedPlanetHabitable}} & \shortstack{Exoplanet Archive Calculated} & \shortstack{\citet{Quintana2014EarthSizedPlanetHabitable} \\ \citet{Torres2015Validation12Small}} & \shortstack{Q1-17 KOI Table \\ \citet{Torres2015Validation12Small}} & \shortstack{\citet{Quintana2014EarthSizedPlanetHabitable} \\ \citet{Torres2015Validation12Small}} \\
Kepler-1987 & \shortstack{Q1-17 KOI Table} & \shortstack{Exoplanet Archive Calculated} & \shortstack{Q1-17 KOI Table} & \shortstack{Q1-17 KOI Table} & \shortstack{Q1-17 KOI Table}  \\
Kepler-208 & \shortstack{\citet{Morton2016FalsePositiveProbabilities}} & \shortstack{Exoplanet Archive Calculated} & \shortstack{\citet{Rowe2014ValidationKeplersMultiple}} & \shortstack{Q1-17 KOI Table} & \shortstack{\citet{Rowe2014ValidationKeplersMultiple}}  \\
Kepler-215 & \shortstack{\citet{Morton2016FalsePositiveProbabilities}} & \shortstack{Exoplanet Archive Calculated \\ \citet{Hadden2014DensitiesEccentricities139}} & \shortstack{\citet{Rowe2014ValidationKeplersMultiple}} & \shortstack{Q1-17 KOI Table} & \shortstack{\citet{Rowe2014ValidationKeplersMultiple}}  \\
Kepler-220 & \shortstack{\citet{Weiss2024KeplerGiantPlanet}} & \shortstack{Exoplanet Archive Calculated} & \shortstack{\citet{Rowe2014ValidationKeplersMultiple}} & \shortstack{\citet{Weiss2024KeplerGiantPlanet}} & \shortstack{\citet{Rowe2014ValidationKeplersMultiple}}  \\
Kepler-221 & \shortstack{Q1-8 KOI Table} & \shortstack{Exoplanet Archive Calculated \\ \citet{Hadden2014DensitiesEccentricities139}} & \shortstack{\citet{Rowe2014ValidationKeplersMultiple}} & \shortstack{Q1-17 KOI Table} & \shortstack{\citet{Rowe2014ValidationKeplersMultiple}}  \\
Kepler-223 & \shortstack{\citet{Mills2016ResonantChainFour}} & \shortstack{\citet{Mills2016ResonantChainFour}} & \shortstack{\citet{Mills2016ResonantChainFour}} & \shortstack{\citet{Mills2016ResonantChainFour}} & \shortstack{Q1-17 KOI Table} \\
Kepler-224 & \shortstack{Q1-8 KOI Table} & \shortstack{Exoplanet Archive Calculated} & \shortstack{\citet{Rowe2014ValidationKeplersMultiple}} & \shortstack{Q1-17 KOI Table} & \shortstack{\citet{Rowe2014ValidationKeplersMultiple}} \\
Kepler-235 & \shortstack{Q1-8 KOI Table} & \shortstack{Exoplanet Archive Calculated} & \shortstack{\citet{Rowe2014ValidationKeplersMultiple}} & \shortstack{Q1-17 KOI Table} & \shortstack{\citet{Rowe2014ValidationKeplersMultiple}} \\
Kepler-238 & \shortstack{\citet{Morton2016FalsePositiveProbabilities}} & \shortstack{Exoplanet Archive Calculated \\ \citet{Xie2014TransitTimingVariation}} & \shortstack{\citet{Rowe2014ValidationKeplersMultiple} \\ \citet{Xie2014TransitTimingVariation}} & \shortstack{Q1-17 KOI Table} & \shortstack{Q1-17 KOI Table \\ \citet{Rowe2014ValidationKeplersMultiple}}  \\
Kepler-24 & \shortstack{\citet{Ford2012TransitTimingObservations}} & \shortstack{Exoplanet Archive Calculated \\ \citet{Ford2012TransitTimingObservations}} & \shortstack{\citet{Ford2012TransitTimingObservations} \\ \citet{Rowe2014ValidationKeplersMultiple}} & \shortstack{Q1-17 KOI Table} & \shortstack{\citet{Ford2012TransitTimingObservations} \\ \citet{Rowe2014ValidationKeplersMultiple}} \\
Kepler-245 & \shortstack{\citet{Morton2016FalsePositiveProbabilities}} & \shortstack{Exoplanet Archive Calculated \\ \citet{Hadden2014DensitiesEccentricities139}} & \shortstack{\citet{Morton2016FalsePositiveProbabilities} \\ \citet{Rowe2014ValidationKeplersMultiple}} & \shortstack{Q1-17 KOI Table} & \shortstack{Q1-17 KOI Table \\ \citet{Rowe2014ValidationKeplersMultiple}}  \\
Kepler-251 & \shortstack{\citet{Morton2016FalsePositiveProbabilities}} & \shortstack{Exoplanet Archive Calculated} & \shortstack{\citet{Rowe2014ValidationKeplersMultiple}} & \shortstack{Q1-17 KOI Table} & \shortstack{\citet{Rowe2014ValidationKeplersMultiple}}  \\
Kepler-256 & \shortstack{\citet{Morton2016FalsePositiveProbabilities}} & \shortstack{Exoplanet Archive Calculated} & \shortstack{\citet{Rowe2014ValidationKeplersMultiple}} & \shortstack{Q1-17 KOI Table} & \shortstack{\citet{Rowe2014ValidationKeplersMultiple}} \\
Kepler-26 & \shortstack{\citet{Jontof-Hutter2016SecureMassMeasurements}} & \shortstack{Exoplanet Archive Calculated \\ \citet{Jontof-Hutter2016SecureMassMeasurements}} & \shortstack{\citet{Jontof-Hutter2016SecureMassMeasurements} \\ \citet{Rowe2014ValidationKeplersMultiple}} & \shortstack{Q1-17 KOI Table} & \shortstack{\citet{Muirhead2012CharacterizingCoolKepler} \\ \citet{Rowe2014ValidationKeplersMultiple}}  \\
Kepler-265 & \shortstack{\citet{Morton2016FalsePositiveProbabilities}} & \shortstack{Exoplanet Archive Calculated} & \shortstack{\citet{Rowe2014ValidationKeplersMultiple}} & \shortstack{Q1-17 KOI Table} & \shortstack{\citet{Rowe2014ValidationKeplersMultiple}}  \\
Kepler-282 & \shortstack{Q1-17 KOI Table} & \shortstack{Exoplanet Archive Calculated \\ \citet{Xie2014TransitTimingVariation}} & \shortstack{\citet{Rowe2014ValidationKeplersMultiple} \\ \citet{Xie2014TransitTimingVariation}} & \shortstack{Q1-17 KOI Table} & \shortstack{Q1-17 KOI Table \\ \citet{Rowe2014ValidationKeplersMultiple}} \\
Kepler-286 & \shortstack{\citet{Morton2016FalsePositiveProbabilities}} & \shortstack{Exoplanet Archive Calculated} & \shortstack{\citet{Rowe2014ValidationKeplersMultiple}} & \shortstack{Q1-17 KOI Table} & \shortstack{\citet{Rowe2014ValidationKeplersMultiple}}  \\
Kepler-292 & \shortstack{\citet{Morton2016FalsePositiveProbabilities}} & \shortstack{Exoplanet Archive Calculated} & \shortstack{\citet{Rowe2014ValidationKeplersMultiple}} & \shortstack{Q1-17 KOI Table} & \shortstack{\citet{Rowe2014ValidationKeplersMultiple}}  \\
Kepler-299 & \shortstack{\citet{Morton2016FalsePositiveProbabilities}} & \shortstack{Exoplanet Archive Calculated} & \shortstack{\citet{Rowe2014ValidationKeplersMultiple}} & \shortstack{Q1-17 KOI Table} & \shortstack{\citet{Rowe2014ValidationKeplersMultiple}}  \\
Kepler-304 & \shortstack{\citet{Morton2016FalsePositiveProbabilities}} & \shortstack{Exoplanet Archive Calculated} & \shortstack{\citet{Morton2016FalsePositiveProbabilities} \\ \citet{Rowe2014ValidationKeplersMultiple}} & \shortstack{Q1-17 KOI Table} & \shortstack{Q1-17 KOI Table \\ \citet{Rowe2014ValidationKeplersMultiple}}  \\
Kepler-305 & \shortstack{\citet{Xie2014TransitTimingVariation}} & \shortstack{Exoplanet Archive Calculated \\ \citet{Hadden2017KeplerPlanetMasses} \\ \citet{Xie2014TransitTimingVariation}} & \shortstack{\citet{Armstrong2021ExoplanetValidationMachine} \\ \citet{Rowe2014ValidationKeplersMultiple} \\ \citet{Xie2014TransitTimingVariation}} & \shortstack{Q1-17 KOI Table} & \shortstack{Q1-17 KOI Table \\ \citet{Rowe2014ValidationKeplersMultiple}} \\
Kepler-306 & \shortstack{\citet{Morton2016FalsePositiveProbabilities}} & \shortstack{Exoplanet Archive Calculated} & \shortstack{\citet{Rowe2014ValidationKeplersMultiple}} & \shortstack{Q1-17 KOI Table} & \shortstack{\citet{Rowe2014ValidationKeplersMultiple}} \\
Kepler-32 & \shortstack{\citet{Fabrycky2012TransitTimingObservations}} & \shortstack{Exoplanet Archive Calculated \\ \citet{Fabrycky2012TransitTimingObservations}} & \shortstack{\citet{Fabrycky2012TransitTimingObservations}} & \shortstack{Q1-17 KOI Table} & \shortstack{\citet{Fabrycky2012TransitTimingObservations}}  \\
Kepler-324 & \shortstack{\citet{Jontof-Hutter2021FollowingKeplerField}} & \shortstack{Exoplanet Archive Calculated \\ \citet{Jontof-Hutter2021FollowingKeplerField}} & \shortstack{\citet{Jontof-Hutter2021FollowingKeplerField} \\ \citet{Rowe2014ValidationKeplersMultiple} \\ \citet{Valizadegan2022ExoMinerHighlyAccurate}} & \shortstack{Q1-17 KOI Table} & \shortstack{Q1-17 KOI Table \\ \citet{Rowe2014ValidationKeplersMultiple}} \\
Kepler-33 & \shortstack{\citet{Lissauer2012AlmostAllKeplers}} & \shortstack{Exoplanet Archive Calculated \\ \citet{Hadden2016NumericalAnalyticalModeling} \\ \citet{Hadden2017KeplerPlanetMasses}} & \shortstack{\citet{Lissauer2012AlmostAllKeplers}} & \shortstack{\citet{Lissauer2012AlmostAllKeplers}} & \shortstack{\citet{Lissauer2012AlmostAllKeplers}} \\
Kepler-338 & \shortstack{\citet{Weiss2024KeplerGiantPlanet}} & \shortstack{Exoplanet Archive Calculated \\ \citet{Hadden2014DensitiesEccentricities139}} & \shortstack{\citet{Hadden2014DensitiesEccentricities139} \\ \citet{Rowe2014ValidationKeplersMultiple}} & \shortstack{\citet{VanEylen2015EccentricityTransitPhotometry}} & \shortstack{\citet{Rowe2014ValidationKeplersMultiple} \\ \citet{Weiss2024KeplerGiantPlanet}}  \\
Kepler-341 & \shortstack{\citet{Morton2016FalsePositiveProbabilities}} & \shortstack{Exoplanet Archive Calculated} & \shortstack{\citet{Rowe2014ValidationKeplersMultiple}} & \shortstack{Q1-17 KOI Table} & \shortstack{\citet{Rowe2014ValidationKeplersMultiple}}  \\
Kepler-342 & \shortstack{\citet{Morton2016FalsePositiveProbabilities}} & \shortstack{Exoplanet Archive Calculated} & \shortstack{\citet{Morton2016FalsePositiveProbabilities} \\ \citet{Rowe2014ValidationKeplersMultiple}} & \shortstack{Q1-17 KOI Table} & \shortstack{Q1-17 KOI Table \\ \citet{Rowe2014ValidationKeplersMultiple}}  \\
Kepler-37 & \shortstack{\citet{Bonomo2023ColdJupitersImproved}} & \shortstack{\citet{Bonomo2023ColdJupitersImproved} \\ \citet{Weiss2024KeplerGiantPlanet}} & \shortstack{\citet{Bonomo2023ColdJupitersImproved} \\ Q1-8 KOI Table} & \shortstack{\citet{Bonomo2023ColdJupitersImproved} \\ \citet{Weiss2024KeplerGiantPlanet}} & \shortstack{\citet{Bonomo2023ColdJupitersImproved} \\ \citet{Weiss2024KeplerGiantPlanet}}  \\
Kepler-402 & \shortstack{\citet{Morton2016FalsePositiveProbabilities}} & \shortstack{Exoplanet Archive Calculated} & \shortstack{\citet{Rowe2014ValidationKeplersMultiple}} & \shortstack{Q1-17 KOI Table} & \shortstack{\citet{Rowe2014ValidationKeplersMultiple}} \\
Kepler-48 & \shortstack{\citet{Marcy2014MassesRadiiOrbits}} & \shortstack{\citet{Marcy2014MassesRadiiOrbits}} & \shortstack{\citet{Marcy2014MassesRadiiOrbits}} & \shortstack{\citet{Weiss2024KeplerGiantPlanet}} & \shortstack{\citet{Weiss2024KeplerGiantPlanet}} \\
Kepler-49 & \shortstack{\citet{Steffen2013TransitTimingObservations}} & \shortstack{Exoplanet Archive Calculated \\ \citet{Hadden2017KeplerPlanetMasses}} & \shortstack{\citet{Rowe2014ValidationKeplersMultiple} \\ \citet{Steffen2013TransitTimingObservations}} & \shortstack{Q1-17 KOI Table} & \shortstack{\citet{Muirhead2012CharacterizingCoolKepler} \\ \citet{Rowe2014ValidationKeplersMultiple}}  \\
Kepler-55 & \shortstack{\citet{Steffen2013TransitTimingObservations}} & \shortstack{Exoplanet Archive Calculated \\ \citet{Hadden2014DensitiesEccentricities139}} & \shortstack{\citet{Rowe2014ValidationKeplersMultiple} \\ \citet{Steffen2013TransitTimingObservations}} & \shortstack{Q1-17 KOI Table} & \shortstack{Q1-17 KOI Table \\ \citet{Rowe2014ValidationKeplersMultiple}}  \\
Kepler-58 & \shortstack{\citet{Steffen2013TransitTimingObservations}} & \shortstack{Exoplanet Archive Calculated \\ \citet{Steffen2013TransitTimingObservations}} & \shortstack{Q1-17 KOI Table \\ \citet{Rowe2014ValidationKeplersMultiple} \\ \citet{Steffen2013TransitTimingObservations}} & \shortstack{Q1-17 KOI Table} & \shortstack{Q1-17 KOI Table \\ \citet{Rowe2014ValidationKeplersMultiple}} \\
Kepler-62 & \shortstack{\citet{Borucki2013Kepler62FivePlanetSystem}} & \shortstack{\citet{Borucki2013Kepler62FivePlanetSystem}} & \shortstack{\citet{Borucki2013Kepler62FivePlanetSystem}} & \shortstack{\citet{Weiss2024KeplerGiantPlanet}} & \shortstack{\citet{Borucki2013Kepler62FivePlanetSystem}}  \\
Kepler-65 & \shortstack{\citet{Chaplin2013AsteroseismicDeterminationObliquities}} & \shortstack{\citet{Mills2019LongperiodGiantCompanions}} & \shortstack{\citet{Chaplin2013AsteroseismicDeterminationObliquities}} & \shortstack{\citet{Mills2019LongperiodGiantCompanions}} & \shortstack{\citet{Chaplin2013AsteroseismicDeterminationObliquities}}  \\
Kepler-758 & \shortstack{\citet{Morton2016FalsePositiveProbabilities}} & \shortstack{Exoplanet Archive Calculated} & \shortstack{\citet{Morton2016FalsePositiveProbabilities}} & \shortstack{Q1-17 KOI Table} & \shortstack{Q1-17 KOI Table}  \\
Kepler-79 & \shortstack{\citet{Jontof-Hutter2014Kepler79sLowDensity}} & \shortstack{\citet{Jontof-Hutter2014Kepler79sLowDensity}} & \shortstack{\citet{Jontof-Hutter2014Kepler79sLowDensity}} & \shortstack{\citet{Jontof-Hutter2014Kepler79sLowDensity}} & \shortstack{\citet{Jontof-Hutter2014Kepler79sLowDensity}} \\
Kepler-80 & \shortstack{\citet{MacDonald2016DynamicalAnalysisKepler80}} & \shortstack{Exoplanet Archive Calculated \\ \citet{MacDonald2016DynamicalAnalysisKepler80}} & \shortstack{\citet{MacDonald2016DynamicalAnalysisKepler80} \\ \citet{Shallue2018IdentifyingExoplanetsDeep}} & \shortstack{N/A \\ Q1-17 KOI Table} & \shortstack{\citet{MacDonald2016DynamicalAnalysisKepler80} \\ N/A}  \\
Kepler-82 & \shortstack{\citet{Freudenthal2019KeplerObjectInterest}} & \shortstack{Exoplanet Archive Calculated \\ \citet{Freudenthal2019KeplerObjectInterest}} & \shortstack{\citet{Freudenthal2019KeplerObjectInterest} \\ \citet{Rowe2014ValidationKeplersMultiple}} & \shortstack{\citet{Freudenthal2019KeplerObjectInterest} \\ Q1-17 KOI Table} & \shortstack{\citet{Freudenthal2019KeplerObjectInterest} \\ \citet{Rowe2014ValidationKeplersMultiple}} \\
Kepler-84 & \shortstack{\citet{Xie2013TransitTimingVariation}} & \shortstack{Exoplanet Archive Calculated \\ \citet{Hadden2014DensitiesEccentricities139}} & \shortstack{\citet{Rowe2014ValidationKeplersMultiple} \\ \citet{Xie2013TransitTimingVariation}} & \shortstack{Q1-17 KOI Table} & \shortstack{Q1-17 KOI Table \\ \citet{Rowe2014ValidationKeplersMultiple}} \\
Kepler-85 & \shortstack{\citet{Xie2013TransitTimingVariation}} & \shortstack{Exoplanet Archive Calculated \\ \citet{Hadden2014DensitiesEccentricities139} \\ \citet{Hadden2017KeplerPlanetMasses}} & \shortstack{\citet{Rowe2014ValidationKeplersMultiple} \\ \citet{Xie2013TransitTimingVariation}} & \shortstack{Q1-17 KOI Table} & \shortstack{Q1-17 KOI Table \\ \citet{Rowe2014ValidationKeplersMultiple}}  \\
L 98-59 & \shortstack{\citet{Demangeon2021WarmTerrestrialPlanet}} & \shortstack{\citet{Demangeon2021WarmTerrestrialPlanet} \\ \citet{Luque2022DensityNotRadius}} & \shortstack{\citet{Demangeon2021WarmTerrestrialPlanet} \\ \citet{Luque2022DensityNotRadius}} & \shortstack{\citet{Demangeon2021WarmTerrestrialPlanet} \\ \citet{Luque2022DensityNotRadius}} & \shortstack{\citet{Cloutier2019Characterization9859Multiplanetary} \\ \citet{Demangeon2021WarmTerrestrialPlanet}} \\
TOI-1136 & \shortstack{\citet{Dai2023TOI1136YoungCoplanar}} & \shortstack{\citet{Beard2024TESSKeckSurveyXVII}} & \shortstack{\citet{Dai2023TOI1136YoungCoplanar}} & \shortstack{\citet{Beard2024TESSKeckSurveyXVII}} & \shortstack{\citet{Beard2024TESSKeckSurveyXVII} \\ N/A}  \\
TOI-1246 & \shortstack{\citet{Turtelboom2022TESSKeckSurveyXI}} & \shortstack{\citet{Turtelboom2022TESSKeckSurveyXI}} & \shortstack{\citet{Turtelboom2022TESSKeckSurveyXI}} & \shortstack{N/A} & \shortstack{\citet{Turtelboom2022TESSKeckSurveyXI}} \\
TOI-178 & \shortstack{\citet{Leleu2021SixTransitingPlanets}} & \shortstack{\citet{Leleu2021SixTransitingPlanets}} & \shortstack{\citet{Leleu2021SixTransitingPlanets}} & \shortstack{N/A} & \shortstack{\citet{Leleu2021SixTransitingPlanets}} \\
TOI-4010 & \shortstack{\citet{Kunimoto2023TOI4010SystemThree}} & \shortstack{\citet{Kunimoto2023TOI4010SystemThree}} & \shortstack{\citet{Kunimoto2023TOI4010SystemThree}} & \shortstack{\citet{Kunimoto2023TOI4010SystemThree}} & \shortstack{\citet{Kunimoto2023TOI4010SystemThree}} \\
TOI-500 & \shortstack{\citet{Serrano2022LoweccentricityMigrationPathway}} & \shortstack{\citet{Serrano2022LoweccentricityMigrationPathway}} & \shortstack{\citet{Serrano2022LoweccentricityMigrationPathway}} & \shortstack{\citet{Serrano2022LoweccentricityMigrationPathway}} & \shortstack{\citet{Serrano2022LoweccentricityMigrationPathway}}  \\
TOI-561 & \shortstack{\citet{Lacedelli2022InvestigatingArchitectureInternal}} & \shortstack{\citet{Lacedelli2022InvestigatingArchitectureInternal}} & \shortstack{\citet{Lacedelli2022InvestigatingArchitectureInternal}} & \shortstack{\citet{Lacedelli2022InvestigatingArchitectureInternal}} & \shortstack{\citet{Lacedelli2022InvestigatingArchitectureInternal}}  \\
TOI-700 & \shortstack{\citet{Gilbert2023SecondEarthsizedPlanet}} & \shortstack{Exoplanet Archive Calculated} & \shortstack{\citet{Gilbert2023SecondEarthsizedPlanet}} & \shortstack{\citet{Gilbert2023SecondEarthsizedPlanet}} & \shortstack{\citet{Gilbert2023SecondEarthsizedPlanet}}  \\
TRAPPIST-1 & \shortstack{\citet{Agol2021RefiningTransittimingPhotometric}} & \shortstack{\citet{Agol2021RefiningTransittimingPhotometric}} & \shortstack{\citet{Agol2021RefiningTransittimingPhotometric}} & \shortstack{\citet{Grimm2018NatureTRAPPIST1Exoplanets}} & \shortstack{\citet{Agol2021RefiningTransittimingPhotometric}}  \\
WASP-47 & \shortstack{\citet{Nascimbeni2023NewDynamicalModeling}} & \shortstack{\citet{Nascimbeni2023NewDynamicalModeling}} & \shortstack{\citet{Nascimbeni2023NewDynamicalModeling}} & \shortstack{\citet{Nascimbeni2023NewDynamicalModeling}} & \shortstack{\citet{Crossfield2016197Candidates104} \\ \citet{Hellier2012SevenTransitingHot}} 
\enddata

\end{deluxetable*}

\end{longrotatetable}

\bibliography{references,references-3,osg}
\bibliographystyle{aasjournalv7}

\end{document}